\documentclass[prl,showpacs,twocolumn]{revtex4}

\newcommand {\etal}{{\it et al.}}         
\newcommand {\rut}{\rho _{\rm{ut}}}
\newcommand {\rt}{\rho _{\rm{t}}}

\newcommand {\NPIO}{(Nd$_{1-x}$Pr$_{x}$)$_2$Ir$_2$O$_7$ }
\newcommand {\SNIO}{(Sm$_{y}$Nd$_{1-y}$)$_2$Ir$_2$O$_7$ }

\usepackage[dvipdfmx]{graphicx}
\usepackage{bm}
\usepackage{pifont}
\usepackage{amsmath,amssymb}
\usepackage[dvipdfmx]{color} 

\begin{document}

\title{Pressure and magnetic-field effects on metal-insulator transitions of bulk and domain-wall states in pyrochlore iridates}

\author{K. Ueda$^1$}
\author{J. Fujioka$^{1}$}
\author{C. Terakura$^{2}$}
\author{Y. Tokura$^{1,2}$}

\affiliation{
$^1$ Department of Applied Physics, University of Tokyo, Tokyo 113-8656, Japan\\
$^2$ RIKEN Center for Emergent Matter Science (CEMS), Wako 351-0198, Japan\\
}

\date{July 17, 2015}

\begin{abstract}
We have explored the critical metal-insulator phenomena for pyrochlore-type $R_2$Ir$_2$O$_7$, in which electron correlation strength and magnetic configuration are systematically controlled by varying the average rare-earth ionic radius ($R$=Nd$_{1-x}$Pr$_{x}$ and Sm$_{y}$Nd$_{1-y}$), external pressure, and magnetic field.
Metal-insulator transitions in bulk are caused by increasing $x$ or tuning external pressure, indicating that the effective electron correlation is responsible for the transition.
The metallic state intervenes between the paramagnetic insulating and antiferromagnetically ordered insulating phases for \SNIO ($y$=0.7-0.9), reminiscent of the first-order Mott transition.
Furthermore, the metal-to-insulator crossover is observed (around $y$=0.7) for the charge transport on magnetic domain walls in the insulating bulk. 
An application of magnetic field also drives metal-insulator transitions for \NPIO in which a variety of exotic topological quantum states are potentially realized.
\end{abstract}
\maketitle

Strong electron correlation in materials often generates intriguing electronic phase diagrams in which the balance among various competing phases can be tuned through parameters such as chemical doping, pressure, and magnetic field\cite{rev-MIT,rev-CMR}.
Recently, the interplay between effective electron correlation ($U$) and relativistic spin-orbit coupling (SOC) has stimulated considerable interest as an important origin for novel electronic/magnetic phases\cite{Krempa_rev}.
For instance, in a family of $5d$ oxides where the energy scale of $U$ is comparable to the SOC energy and the band width of $J_{\rm{eff}}=1/2$ state, many unconventional properties have been increasingly discovered such as relativistic Mott insulator\cite{BJKimPRL,BJKimscience}, novel spin-orbit exciton\cite{JHKim_SOexciton2012,BHKim_SOexciton,JHKim_SOexciton2014}, and unique magnetisms accompanying Kitaev-type anisotropic interaction\cite{JackeliKhaliullin,Na2IrO3Choi2012,Ohgushi2013,harmonichoneycomb2014}.

Among them, pyrochlore-type $R_{2}$Ir$_{2}$O$_{7}$ ($R$IO), which consist of the corner-linked rare-earth $R$ ions and Ir tetrahedra as shown in Fig. 1 (a)\cite{pyrochlore-rev}, are suggested to host emergent topological quantum states including Weyl semimetal and topological Mott insulator in the vicinity of metal-insulator transition (MIT)\cite{Krempa_rev,NPhysBalents,PRBYang,2011Imada,XWan,2012Krempa,2013Krempa,2013PRLSBLee,PRXYamaji,PRLYang}.
$R$IO provides an ideal platform to study the interplay between $U$ and SOC since we can systematically tune $U$ with changing ionic radius of $R$-ion ($r$).
In reality, $R$IO exhibit MIT by varying $R$ ions; the Pr$_{2}$Ir$_{2}$O$_{7}$, which is characterized by the largest $r$ among the family of $R$IO, is a paramagnetic metal down to 0.12 K\cite{2006PIO}, whereas others with relatively small $R$ ionic radii display thermally-induced MIT concomitant with all-in all-out (AIAO) type antiferromagnetic order (see the middle panel of Fig. 1 (a)) below $T_{\rm{N}}$ or otherwise remain insulating over the whole temperature $T$ region \cite{2011Matsuhira,NIO-neutron,EIO-Xray}.
The Pr$_{2}$Ir$_{2}$O$_{7}$ shows unconventional magnetotransport phenomena possibly reflecting its characteristic band dispersion in $k$-space\cite{2007Machida,2010Machida,2011PIO}.
For the second largest $R$=Nd compound Nd$_{2}$Ir$_{2}$O$_{7}$, unique properties such as anomalous $\omega $-linear optical conductivity spectra\cite{NIO-MIT} and large magnetoresistance\cite{NIO-GMR,2013Disseler,NIO-DW} are discerned below $T_{\rm{N}}$.
In particular, the anomalous metallic conduction is observed on antiferromagnetic domain walls (DWs) embedded in the fully-gapped insulating bulk\cite{NIO-DW}.
Furthermore, the magnetic field-induced transition occurs from the AIAO insulator to the 2-in 2-out (see the right panel of Fig. 1 (a)) semimetal, likely associated with the altered topological nature of Ir-$5d$ band structure\cite{NIO-single}.
These findings indicate that the cooperation between $U$ and SOC on the boundary of MIT in $R$IO may potentially give rise to abundant emergent quantum states; to unravel them it is necessary to control $U$ more finely and precisely in the MIT critical region than in previous studies.


In this study, we have investigated the transport properties for well-amalgamated polycrystals of \SNIO and \NPIO combined with application of external pressure and magnetic field.
Application of external pressure is a well-known way to tune $U$ without introducing disorders.
On the other hand, the chemical substitution enables us not only to control $U$ quite precisely (``chemical" pressure effect) but also to employ various experimental techniques including magnetization and magneto-transport measurements.
Indeed, the relationship between the one-electron band width and the average size ($r$) of rare-earth ions has been well established not only for perovskites\cite{rev-MIT,rev-CMR}, but also for pyrochlores\cite{pyrochlore-rev,2009Mo_Iguchi,2011Matsuhira}.
Moreover, the magnetic characters of Pr-$4f$ electrons are similar to those of Nd-$4f$ ones; the Pr(Nd)-$4f$ magnetic moment is $\sim 2.68 \mu_{\rm{B}}$/Pr\cite{PIO-neutron} ($\sim 2.37 \mu_{\rm{B}}$/Nd\cite{NIO-neutron}) and the $f$-$d$ exchange interaction energy are comparable for $R$=Pr\cite{2006PIO} and $R$=Nd\cite{NIO-neutron,NIO-MIT}.
Therefore, the present mixed-crystal systems provide an ideal system to study the magnetic properties as well as the band-width control effect.
We have found that the compounds of $y$=0.7-0.9 exhibit the insulator-metal-insulator transition reminiscent of first-order Mott transition.
Around there, the metal-to-insulator crossover on magnetic DWs is discerned, implying that the electronic state on DWs are intimately related to the bulk state.
In addition, large magnetoresistance is observed for $x>$0.4, attributable to magnetic field-induced phase transitions among a variety of topological quantum states.


\begin{figure}
\begin{center}
\includegraphics[width=3.2in,keepaspectratio=true]{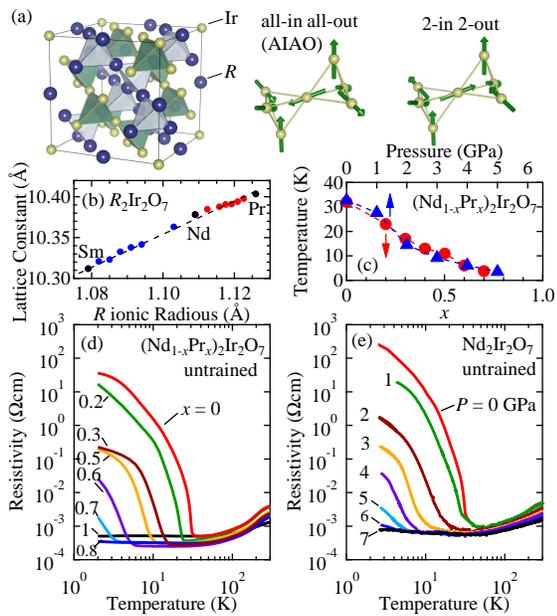}
\caption{(color online) 
(a) Pyrochlore lattice structure composed of Ir and rare-earth ion ($R$) tetrahedrons (left panel), and the all-in all-out (middle panel) and the 2-in 2-out (right panel) magnetic ordered configurations.
(b) Lattice constant as a function of $R$ ionic radius for (Sm$_{y}$Nd$_{1-y}$)$_2$Ir$_2$O$_7$ and (Nd$_{1-x}$Pr$_{x}$)$_2$Ir$_2$O$_7$.
(c) Hydrostatic pressure dependence of magnetic ordering or equivalently the metal-insulator transition temperature for Nd$_2$Ir$_2$O$_7$ ($x$=0) in comparison with the chemical pressure dependence via varying the composition $x$ in (Nd$_{1-x}$Pr$_{x}$)$_2$Ir$_2$O$_7$.
(d) Temperature dependence of resistivity for several compounds of (Nd$_{1-x}$Pr$_{x}$)$_2$Ir$_2$O$_7$.
(e) Temperature dependence of resistivity for $R$=Nd at several hydrostatic pressures.
All resistivity curves in (d) and (e) were measured after zero-field cooling (untrained state).
}

\end{center}
\end{figure}

The polycrystalline samples (Sm$_{y}$Nd$_{1-y}$)$_2$Ir$_2$O$_7$ and (Nd$_{1-x}$Pr$_{x}$)$_2$Ir$_2$O$_7$ were prepared by solid-state high-pressure synthesis as described in Ref.\cite{NIO-MIT}.
The high-pressure synthesis, while keeping the stoichiometry crucial for $R$IO\cite{EIOnonst}, provides the high-quality samples which allows us to study the systematic $R$-substitution effect.
As shown in Fig. 1 (b), the change of the lattice constant with varying $r$ or chemical pressure follows the Vegard's law.

Figure 1 (d) exhibits the temperature ($T$) dependence of resistivity for several compounds of (Nd$_{1-x}$Pr$_{x}$)$_2$Ir$_2$O$_7$, all of which were measured under zero magnetic field (untrained state).
Although the observed resistivity at low $T$ below $T_{\rm{N}}$ includes the contributions from both DWs and bulk itself, we can discuss the properties of the bulk because the resistivity of untrained state is nearly scaled with the bulk resistivity (trained state)\cite{NIO-DW}.
The insulating phase below $T_{\rm{N}}$ is systematically suppressed with increasing the Pr-ion population $x$, and the transition disappears around $x$=0.8 with no resistivity upturn down to 2 K.
We also measured the $T$ dependence of resistivity for Nd$_2$Ir$_2$O$_7$ under several hydrostatic pressures as shown in Fig. 1 (e) by employing the cubic anvil cell\cite{cubicanvil}.
Similarly to the chemical pressure effect, the sharp increase of resistivity below $T_{\rm{N}}$ is gradually suppressed as pressure increases, and the transition is no longer observable down to 3 K beyond 7 GPa, in accord with the reported result in ref.\cite{NIO_pressure}.
To compare the effect between chemical substitution and external pressure, we plot the $T_{\rm{N}}$ which can be determined by the sharp increase of resistivity\cite{EIO_pressure} as a function of pressure and $x$ in Fig. 1 (c).
The $T_{\rm{N}}$ appears to decrease monotonically with increasing pressure ($\Delta P$) and increasing $x$ ($\Delta x$), well scaled to each other; $\Delta P\sim 0.7$ GPa corresponds to $\Delta x\sim 0.1$.
We note that the physical pressure is somewhat different from the chemical pressure; the larger $R$ atomic size increases both the Ir-O-Ir bond angle and length\cite{2001Mo_Morimoto,Millican} whereas the pressure substantially decreases the latter\cite{2004Mo_P}.
One might expect that these differences should induce the different modulation of some parameters such as direct Ir-Ir hopping\cite{2012Krempa,2013Krempa}.
Nevertheless, the observed resistivity curve with changing chemical substitution shows a well parallel behavior with that with tuning external pressure; thus the $U$ may be responsible for the electronic transport properties in the range of $r$ discussed in the present paper.

\begin{figure}
\begin{center}
\includegraphics[width=3.4in,keepaspectratio=true]{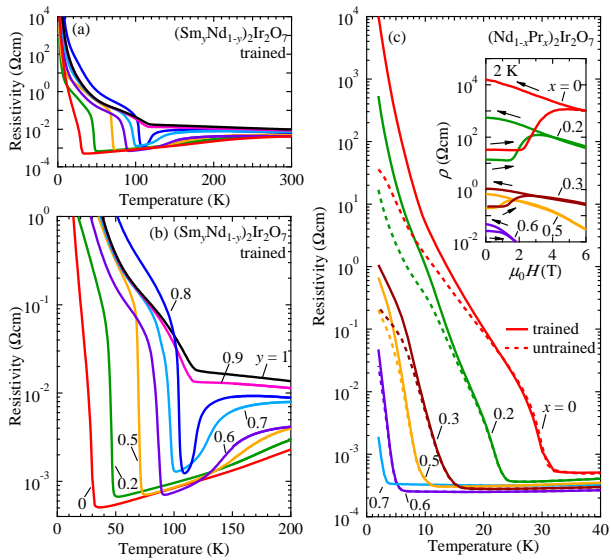}
\caption{(color online). 
(a) Temperature dependence of resistivity for (Sm$_{y}$Nd$_{1-y}$)$_2$Ir$_2$O$_7$, and (b) its enlarged view in the temperature range from 0 K to 200 K.
(c) Temperature dependence of resistivity for (Nd$_{1-x}$Pr$_{x}$)$_2$Ir$_2$O$_7$. The solid lines are resistivity of trained state measured at 0 T on warming run after 14 T-field cooling and the broken ones are that of untrained state measured after zero-field cooling.
The inset shows the magnetic field dependence of resistivity for several compositions. Starting from the zero field-cooled (untrained) state, these curves were measured for increasing and decreasing magnetic field scans as shown by arrows.
}
\end{center}
\end{figure}

In this context, $R$-solid-solution compounds are quite suitable for systematic study of bandwidth control effect on $R_2$Ir$_2$O$_7$.
The $T$ dependence of resistivity of trained state (14 T-field cooled) for \SNIO ($0<y<1$) are presented in Figs. 2 (a) and 2 (b).
At high $T$, the resistivity for $y$=0-0.6 decreases with lowering $T$ ($d\rho /dT>0$), whereas it shows rather thermally-activated behavior ($d\rho /dT<0$) for $y$=0.9-1.
The resistivity for each $y$ significantly increases below $T_{\rm{N}}$ which systematically shifts to higher $T$ as $y$ increases.
Nonmonotonous $T$ dependence of resistivity is discernible especially for $y$=0.8.
The resistivity slightly increases with lowering $T$ in a high $T$ region, then abruptly decreases by one order of magnitude below 150 K, and eventually diverges accompanying AIAO magnetic order below $T_{\rm{N}}$=106 K.
This is reminiscent of the transitions among a paramagnetic Mott insulator, a correlated metal and an antiferromagnetic insulator as sometimes observed in strongly correlated systems\cite{2003V2O3}, although the clear first-order nature accompanying $T$ hysteresis is not discerned apart from the sharp change of resistivity in the present system.
Since the similar reduction of resistivity in an intermediate $T$ region was also observed in the study on pressure effect for $R$=Eu\cite{EIO_pressure}, the observed paramagnetic insulator-metal transition can be attributed neither to the increased disorder nor to the phase separation.

Figure 2 (c) displays the $T$ dependence of resistivity for (Nd$_{1-x}$Pr$_{x}$)$_2$Ir$_2$O$_7$.
The resistivity for $x$=0-0.7 conspicuously increases below respective $T_{\rm{N}}$.
Importantly, all the AIAO insulators of (Nd$_{1-x}$Pr$_{x}$)$_2$Ir$_2$O$_7$ show the difference of resistivity between the trained (14 T-magnetic-field cooled) and untrained (zero-field cooled) states attributable to the existence of metallic state on the AIAO DWs\cite{NIO-DW,NIO-single}.
The realization of metallic DWs is manifested also by the magnetic field dependence of resistivity shown in the inset of Fig. 2 (c); irreversible behaviors of resistivity between field increasing and decreasing scans starting from the untrained states are due to the field alignment of the AIAO-type magnetic domain.
The critical field for such elimination of DWs decreases as $x$ increases, indicating the gradual decline of magnetic anisotropic energy.

\begin{figure}
\begin{center}
\includegraphics[width=3.4in,keepaspectratio=true]{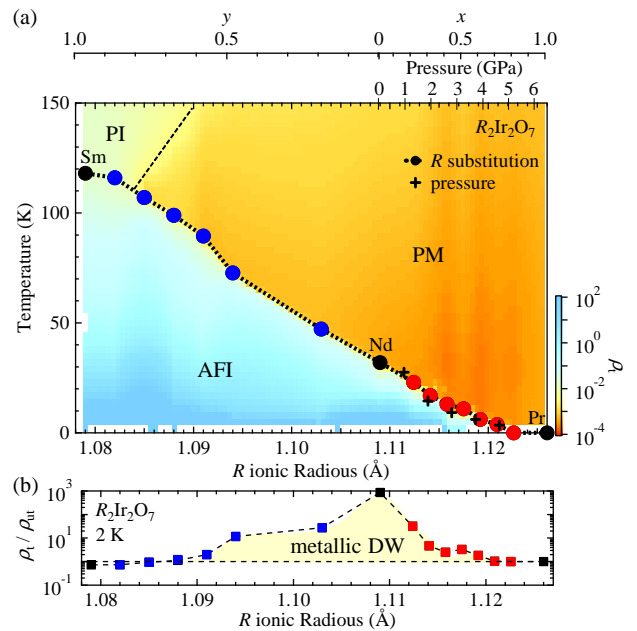}
\caption{(color online). 
(a) Contour map in the plane of $R$ ionic radius ($r$) and temperature for resistivity in (Sm$_{y}$Nd$_{1-y}$)$_2$Ir$_2$O$_7$ and (Nd$_{1-x}$Pr$_{x}$)$_2$Ir$_2$O$_7$ by the interpolation of the experimental data shown in Fig. 2. Dots denote the transition temperature. PM, PI, and AFI stand for the paramagnetic metal, paramagnetic insulator, and antiferromagnetic insulator phase, respectively. The broken line as the border of PI and PM is the guide to the eyes. (b) The ratio of resistivity between trained ($\rt $) and untrained ($\rut $) state as a function of $R$ ionic radius.
}
\end{center}
\end{figure}

The $r$ vs. $T$ phase diagram for (Sm$_{y}$Nd$_{1-y}$)$_2$Ir$_2$O$_7$ and (Nd$_{1-x}$Pr$_{x}$)$_2$Ir$_2$O$_7$ based on the transport results (Figs. 1 and 2) is shown in Fig. 3 (a).
Metallic state with no magnetic order is apparent down to 2 K for large $r$ (Pr-rich regime; $x>0.8$), whereas others exhibit thermal transitions from paramagnetic metal (PM, $d\rho /dT>0$) or paramagnetic insulator (PI, $d\rho /dT<0$) to antiferrromagnetic insulator (AFI) below $T_{\rm{N}}$.
In particular, for (Sm$_{y}$Nd$_{1-y}$)$_2$Ir$_2$O$_7$ ($0.6<y<0.8$) the reentrant insulator-metal-insulator transition is observed as argued above.
$T_{\rm{N}}$ increases rapidly with the decline of $r$ in the range from 1.121 \AA  ($x$=0.7) to 1.079 \AA  ($y$=1).
This indicates that the $T_{\rm{N}}$ is intimately linked to the $U$ which changes almost linearly with varying $r$ in the pyrochlore oxides, being consistent with the theoretical prediction\cite{2013Krempa}.

We have also plotted the ratio of resistivity of trained to untrained state ($\rt /\rut $) as a function of $r$ in Fig. 3 (b).
The $\rt /\rut $ can be regarded as the ratio of conductance between DWs and bulk on the basis of a simple picture of parallel circuit\cite{NIO-DW}.
The $\rt /\rut $ markedly increases with decreasing $r$ and reaches maximum at Nd$_{2}$Ir$_{2}$O$_{7}$ likely due to the smaller value of bulk conductivity.
Subsequently, it decreases significantly as $r$ decreases, implying that the conductance of DWs decreases with increasing $U$.
It is noteworthy that the $\rt /\rut $ becomes less than 1 for $y>0.7$, indicating that the presence of DWs does not assist but rather disrupt the charge current.
In other words, the electronic state on DWs turns from metallic to insulating around $y\sim 0.7$ which is close to the crossover region between PM and PI in the bulk.
To obtain insight about the observed MIT on DWs, it is important to consider the bulk state.
Weyl semimetal with the AIAO magnetic order breaking time-reversal symmetry is predicted to exist in the narrow region right below $T_{\rm{N}}$; each Weyl point approaches rapidly each other in $k$-space and vanishes with slight increase of magnetic order parameter\cite{2012Krempa,2013Krempa,PRXYamaji}.
Once Weyl fermions appear, however, they can imprint metallic Fermi arc states on DWs which may host the observed DW conduction\cite{PRXYamaji}.
In this regard, the observed MIT on DWs around $y\sim $0.7 has important implications for the bulk electronic state; for example, Weyl semimetal state may not show up between PI and AFI for the stronger electron correlation such as $y>$0.7.

\begin{figure}
\begin{center}
\includegraphics[width=3.4in,keepaspectratio=true]{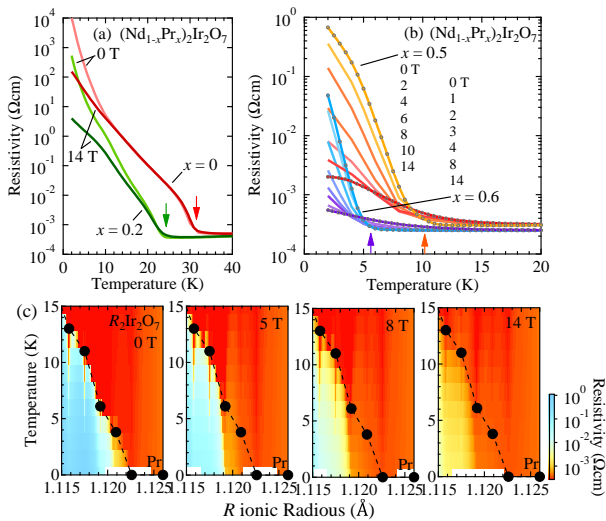}
\caption{(color online).
Temperature dependence of resistivity under several magnetic fields for (a) $x$=0, 0.2 and (b) $x$=0.5, 0.6 in (Nd$_{1-x}$Pr$_{x}$)$_2$Ir$_2$O$_7$. (c) Contour plots of resistivity on the $R$ ionic radius versus temperature plane for several magnetic fields (from left to right: 0, 5, 8, and 14 T). The dots connected with broken lines common to all the figures in (c) represent the metal-insulator phase boundary at zero field.
}
\end{center}
\end{figure}

We now turn to the detailed transport properties of (Nd$_{1-x}$Pr$_{x}$)$_2$Ir$_2$O$_7$ locating near the metal-insulator boundary.
Figures 4 (a) and (b) show the $T$ dependence of resistivity under applied magnetic fields for several compositions.
For Nd-rich compositions ($x$=0, 0.2) in Fig. 4 (a), the magnetoresistance is rather small even in the field of 14 T.
On the other hand, for $x$=0.5 or 0.6 in Fig. 4 (b), the resistivity significantly decreases with increasing field and the anomaly at $T_{\rm{N}}$ completely disappears in a sufficient large field.
The large magnetoresistance is presumably attributed to the MIT with change of the magnetic configuration, as previously shown for single crystal of Nd$_2$Ir$_2$O$_7$ ($x$=0)\cite{NIO-single}.
Figure 4 (c) displays the contour plots of resistivity in the range of $x$=0.4-1.0 for several magnetic fields.
The AFI phase is gradually suppressed and turns into the metallic one with increasing magnetic field for $x>0.4$, whereas the resistivity remains rather insulating for $x<0.3$ as shown in Fig. 4 (a).
According to the predicted phase diagram for the magnetic structural change from AIAO to 2-in 2-out state, as derived by mean-field theory \cite{NIO-single}, the compositions for $x>0.4$ can access to the metallic state with large density of states (DOS) at Fermi level ($E_{\rm{F}}$) by applying magnetic field, whereas others remain in the barely-insulating or semimetallic state with vanishingly-small DOS at $E_{\rm{F}}$.
Interestingly, the calculation demonstrates that there should be a number of novel electronic and magnetic phases such as Weyl metal and nodal semimetal below the critical correlation strength\cite{NIO-single}, that corresponds to $x\sim 0.4$ in the present system.
The observed insulator to metal transition in the range from $x$=0.4 to $x$=0.7 may be consequences of such phase changes.
We note that the present magnetotransport properties are observed in polycrystals composed of microcrystal grains of $\sim $ 5 $\mu $m in size in which the bulk nature with well-stoichiometric composition (i.e. half-filling) is preserved but the crystalline axes are randomized with respect to the magnetic field directions; therefore the transport properties are affected by the mixed contributions from the 3-in 1-out and 2-in 2-out states, each of which may induce Weyl (semi-)metal and nodal semimetal, respectively\cite{NIO-single}.
Further studies on the anisotropic magnetotransport are required to elucidate the electronic ground state separably for each magnetic configuration.

In conclusion, we have investigated the charge transport properties related to the metal-insulator transitions for \SNIO and (Nd$_{1-x}$Pr$_{x}$)$_2$Ir$_2$O$_7$ polycrystals where the effective electron correlation is finely controlled.
The paramagnetic metal-insulator crossover shows up around $y$=0.8 accompanying the insulator-metal-insulator reentrant transition.
Around there, the electronic state on the magnetic domain walls turns from metal to insulator.
Large magnetoresistance is observed for (Nd$_{1-x}$Pr$_{x}$)$_2$Ir$_2$O$_7$ with $0.4<x<0.7$ near the zero-field phase boundary between the antiferromagnetic insulator and the paramagnetic semimetal, pointing to the magnetic field-induced metal-insulator transition, in the course of which novel topological quantum states are expected to emerge.

We acknowledge fruitful discussions with B.-J. Yang, N. Nagaosa, Y. Yamaji, and M. Imada.
This work is supported by the Japan Society for the Promotion of Science through the Funding Program for World-Leading Innovative R$\& $D on Science and Technology (FIRST Program) on 'Quantum Science on Strong Correlation' initiated by the Council for Science and Technology Policy and by JSPS Grant-in-Aid for Scientific Research (No. 80609488 and No. 24224009).


\newpage

\begin{thebibliography}{100}
\bibitem{rev-MIT} M. Imada, A. Fujimori, and Y. Tokura, Rev. Mod. Phys. {\bf 70}, 1039 (1998).
\bibitem{rev-CMR} Y. Tokura, Rep. Prog. Phys. {\bf 69}, 797 (2006).
\bibitem{Krempa_rev} W. Witczak-Krempa, G. Chen, Y. B. Kim, and L. Balents, Annu. Rev. Condens. Matter Phys. {\bf 5}, 57 (2014).
\bibitem{BJKimPRL} B. J. Kim {\etal}, Phys.\ Rev.\ Lett.\ {\bf 101}, 076402 (2008).
\bibitem{BJKimscience} B. J. Kim {\etal}, Science {\bf 323}, 1329 (2009).
\bibitem{JHKim_SOexciton2012} J. Kim {\etal}, Phys.\ Rev.\ Lett.\ {\bf 108}, 177003 (2012).
\bibitem{BHKim_SOexciton} B. H. Kim, G. Khaliullin, and B. I. Min, Phys.\ Rev.\ Lett.\ {\bf 109}, 167205 (2012).
\bibitem{JHKim_SOexciton2014} J. Kim {\etal}, Nat. Commun. {\bf 5}, 4453 (2014).
\bibitem{JackeliKhaliullin} G. Jackeli and G. Khaliullin, Phys.\ Rev.\ Lett.\ {\bf 102}, 017205 (2009).
\bibitem{Na2IrO3Choi2012} S. K. Choi {\etal}, Phys.\ Rev.\ Lett.\ {\bf 108}, 127204 (2012).
\bibitem{Ohgushi2013} K. Ohgushi {\etal}, Phys.\ Rev.\ Lett.\ {\bf 110}, 217212 (2013).
\bibitem{harmonichoneycomb2014} K. A. Modic {\etal}, Nat. Commun. {\bf 5}, 423 (2014).
\bibitem{pyrochlore-rev} J. S. Gardner, M. J. P. Gingras, and J. E. Greedan, Rev. Mod. Phys. {\bf 82}, 53 (2010).
\bibitem{NPhysBalents} D. Pesin and L. Balents, Nature Phys. {\bf 6}, 376 (2010).
\bibitem{PRBYang} B.-J. Yang and Y. B. Kim, Phys. Rev. B {\bf 82}, 085111 (2010).
\bibitem{2011Imada} M. Kurita, Y. Yamaji, and M. Imada, J. Phys. Soc. Jpn. {\bf 80}, 044708 (2011).
\bibitem{XWan} X. Wan, A. M. Turner, A. Vishwanath, S. Y. Savrasov, Phys. Rev. B {\bf 83}, 205101 (2011).
\bibitem{2012Krempa} W. Witczak-Krempa and Y. B. Kim, Phys. Rev. B {\bf 85}, 045124 (2012).
\bibitem{2013Krempa} W. Witczak-Krempa, A. Go, and Y. B. Kim, Phys. Rev. B {\bf 87}, 155101 (2013).
\bibitem{2013PRLSBLee} S.-B. Lee, A. Paramekanti, and Y. B. Kim, Phys. Rev. Lett. {\bf 111}, 196601 (2013).
\bibitem{PRXYamaji} Y. Yamaji and M. Imada, Phys. Rev. X {\bf 4}, 021035 (2014).
\bibitem{PRLYang} B.-J. Yang and N. Nagaosa, Phys. Rev. Lett. {\bf 112}, 246402 (2014).
\bibitem{2006PIO} S. Nakatsuji {\etal}, Phys. Rev. Lett. {\bf 96}, 087204 (2006).
\bibitem{2011Matsuhira} K. Matsuhira {\etal}, J. Phys. Soc. Jpn. {\bf 80}, 094701 (2011).
\bibitem{NIO-neutron}K. Tomiyasu {\etal}, J. Phys. Soc. Jpn. {\bf 81}, 034709 (2012).
\bibitem{EIO-Xray} H. Sagayama {\etal}, Phys. Rev. B {\bf 87}, 100403 (2013).
\bibitem{2007Machida} Y. Machida {\etal}, Phys. Rev. Lett. {\bf 98}, 057203 (2007).
\bibitem{2010Machida} Y. Machida {\etal}, nature {\bf 463}, 210 (2010).
\bibitem{2011PIO} L. Balicas, S. Nakatsuji, Y. Machida, and S. Onoda, Phys. Rev. Lett. {\bf 106}, 217204 (2011).
\bibitem{NIO-MIT} K. Ueda {\etal}, Phys. Rev. Lett. {\bf 109}, 136402 (2012).
\bibitem{NIO-GMR} K. Matsuhira, M. Tokunaga, M. Wakeshima, Y. Hinatsu, and S. Takagi, J. Phys. Soc. Jpn. {\bf 82}, 023706 (2013).
\bibitem{2013Disseler} S. M. Disseler, S. R. Giblin, Chetan Dhital, K. C. Lukas, Stephen D. Wilson, and M. J. Graf, Phys. Rev. B {\bf 87}, 060403 (2013).
\bibitem{NIO-DW} K. Ueda {\etal}, Phys. Rev. B {\bf 89}, 075127 (2014).
\bibitem{NIO-single} K. Ueda {\etal}, arXiv:1506.07336.
\bibitem{2009Mo_Iguchi} S. Iguchi {\etal}, Phys. Rev. Lett. {\bf 102}, 136407 (2009).
\bibitem{EIOnonst} J. J. Ishikawa, E. C. T. O'Farrell, and S. Nakatsuji, Phys. Rev. B {\bf 85}, 245109 (2012).
\bibitem{cubicanvil} N. M$\rm{\hat{o}}$ri, H. Takahashi, and N. Takeshita, High Press. Res. {\bf 24}, 225 (2004).
\bibitem{PIO-neutron}Y. Machida {\etal}, J. Phys. Chem. Solids {\bf 66}, 1435 (2005).
\bibitem{NIO_pressure} M. Sakata {\etal}, Phys. Rev. B {\bf 83}, 041102 (2011).
\bibitem{EIO_pressure} F. F. Tafti, J. J. Ishikawa, A. McCollam, S. Nakatsuji, and S. R. Julian, Phys. Rev. B {\bf 85}, 205104 (2012).
\bibitem{2001Mo_Morimoto} Y. Moritomo {\etal}, Phys. Rev. B {\bf 63}, 144425 (2001).
\bibitem{Millican} J. N. Millican {\etal}, Mater. Res. Bull. {\bf 42}, 928 (2007).
\bibitem{2004Mo_P} H. Ishikawa {\etal}, Phys. Rev. B {\bf 70}, 104103 (2004).
\bibitem{2003V2O3} P. Limelette {\etal}, Science {\bf 203}, 89 (2003); F. Kagawa, K. Miyagawa, and K. Kanoda, Nature {\bf 436}, 534 (2005).



\end{thebibliography}
\end{document}